\documentclass[twocolumn]{aastex631}

\usepackage{newtxtext,newtxmath}
\usepackage{graphicx}	
\usepackage{xcolor}
\usepackage{bm}
\usepackage{anyfontsize}

\shorttitle{X-ray emission from $\epsilon$ Lupi}
\shortauthors{Biswas et al.}

\begin{document}

\title{XMM-Newton Perspective of the Unique Magnetic Binary- $\epsilon$ Lupi}

\correspondingauthor{Ayan Biswas}
\email{ayan.biswas@queensu.ca}

\author[0000-0002-1741-6286]{Ayan Biswas}
\affiliation{Department of Physics, Engineering Physics \& Astronomy, Queen’s University, Kingston, Ontario K7L 3N6, Canada}
\affiliation{Department of Physics \& Space Science, Royal Military College of Canada, PO Box 17000, Station Forces, Kingston, ON K7K 7B4, Canada}

\author[0000-0002-1854-0131]{Gregg A. Wade}
\affiliation{Department of Physics \& Space Science, Royal Military College of Canada, PO Box 17000, Station Forces, Kingston, ON K7K 7B4, Canada}

\author[0000-0002-0844-6563]{Poonam Chandra}
\affiliation{National Radio Astronomy Observatory, 520 Edgemont Road, Charlottesville VA 22903, USA }
\affiliation{National Centre for Radio Astrophysics, Tata Institute of Fundamental Research, Ganeshkhind, Pune-411007, India }

\author[0000-0002-5633-7548]{Veronique Petit}
\affiliation{Department of Physics and Astronomy, Bartol Research Institute, University of Delaware, Newark, DE 19716, USA}

\author[0000-0001-8704-1822]{Barnali Das}
\affiliation{CSIRO, Space and Astronomy, PO Box 1130, Bentley, WA 6151, Australia}

\author{Matthew E. Shultz}
\affiliation{Department of Physics and Astronomy, University of Delaware, Newark, DE 19716, USA}

\begin{abstract}

The $\epsilon$ Lupi A (HD 136504) system stands out among magnetic massive binaries as the only short-period binary system in which both components have detectable magnetic fields. The proximity of the magnetospheres of the components leads to magnetospheric interactions, which are revealed as periodic pulses in the radio light curve of this system. In this work, we aim to investigate the magnetospheric interaction phenomenon in the X-ray domain. We observed this system with the XMM-Newton telescope, covering its orbital period. We observe variable X-ray emission with maximum flux near periastron, showing similarity with radio observations. The X-ray spectra show significantly elevated hard X-ray flux during periastron. We attribute the soft X-ray emission to individual magnetospheres, while the hard X-ray emission is explained by magnetospheric interaction, particularly due to magnetic reconnection. However, unlike in the radio, we do not find any significant short-term X-ray bursts. This exotic system can be an ideal target to study magnetospheric interactions in close binaries with organized magnetospheres.

\end{abstract}


\section{Introduction} \label{sec:intro}

Roughly one in ten OB stars are known to host strong ($\sim$kG order), organized (roughly dipolar) magnetic fields \citep{Wade2014, Morel2015, Grunhut2017}. Such strong, organized fields are capable of channeling stellar wind flows towards the magnetic equator, giving rise to regions of magnetically confined winds (MCW) and creating a co-rotating stellar magnetosphere \citep{Babel1997}. The presence of these dense confined winds leads to additional emission or absorption throughout the electromagnetic spectrum \citep{Andre1988, Linsky1992, Shultz2020, Naze2014, Erba2021, Oksala2015, Leto2021, Das2022b, Das2022}. Although the incidence of detectable magnetic fields is nearly 10\% for isolated hot stars, magnetic fields are very rare in the case of close binaries containing a massive star. In the Binarity and Magnetic Interactions in various classes of Stars (BinaMIcS) survey, \cite{Alecian2015} found that less than 1.5\% of binaries containing OBA stars in a close orbit (orbital period of less than 40 days) have a magnetic hot star. This finding immediately implies a strong connection between magnetism and binarity in such systems. However, the effect of binarity on the origin \citep{Schneider2019} and sustenance \citep{Vidal2019} of magnetic fields are not fully understood. The effect of binary interaction in the context of magnetospheric interaction in massive binaries is also poorly constrained from an observational perspective \citep{Naze2017, Shultz2018}. Several previous studies have focused on the effect of binary interaction in pre-main sequence (PMS) stars \citep{Massi2002, Massi2006, Salter2010, Adams2011, Getman2016}, where excess emission in radio and/or X-rays was found to be due to magnetospheric interaction. We can expect similar behavior in main-sequence hot stars with strong magnetic fields, especially in radio and X-ray bands.

In OB stars, the generation of X-rays is predominantly attributed to embedded wind shocks within their dense, supersonic winds \citep{Lucy1980}. The shock-heated plasma reaches high temperatures ($T > 10^7$ K), which fuels the X-ray emission. In the non-magnetic case, the generated X-rays are soft ($\sim$0.5 keV), and the X-ray luminosity scales with the stellar bolometric luminosity as $L_{\rm x}/L_{\rm bol} \sim 10^{-7}$, especially for O-stars \citep{Chlebowski1989, Naze2014}. The X-ray emission serves as a valuable indicator of stellar mass loss and offers insights into the physics governing stellar winds \citep{Leutenegger2006}.

Notably, in magnetic OB stars, wind characteristics undergo significant modifications compared to their non-magnetic counterparts. In non-magnetic stars, wind shocks originate from collisions of gas going in the same direction with different velocities. These velocity jumps are relatively small compared to the terminal wind velocity. However, in the case of a magnetic star, the magnetosphere confines plasma from the opposite magnetic hemispheres along closed magnetic loops, producing magnetically confined wind shocks (MCWS, \citealt{ud-Doula2014}). In modest-sized magnetospheres, the confined wind does not reach the terminal velocity before the shock. This is different from most of the colliding wind binaries for which the winds from both stars reach terminal velocity before the collision. Thus, the shock velocity in this case is higher than in the non-magnetic case, but lower than in colliding wind binaries, resulting in intermediate-energy X-rays ($<$2 keV; see review by \citealt{Naze2014}).  The X-rays produced in this manner are expected to exhibit rotational modulation due to misaligned rotational and magnetic axes \citep{Townsend2007}, which was observed in the system $\theta^1$ Ori C even before the direct measurement of magnetic fields \citep{Gagne1997}. However, this modulation will be minimal if the star’s rotation and magnetic axes are aligned.

This scenario regarding the production of X-rays from solitary magnetic massive stars has become well-established in the past few years via detailed X-ray observations supplemented with simulations \citep{Pittard2010, Oskinova2011, ud-Doula2014}. However, in a survey of X-ray emission from magnetic massive stars by \cite{Naze2014}, some stars are found to deviate from the MCWS scenario. This excess emission may be explained by centrifugal breakout (CBO) events caused by the continuous ejection of magnetically confined plasma \citep{Shultz2020, Owocki2022}.

\begin{figure}
    \centering
    \includegraphics[width=0.47\textwidth]{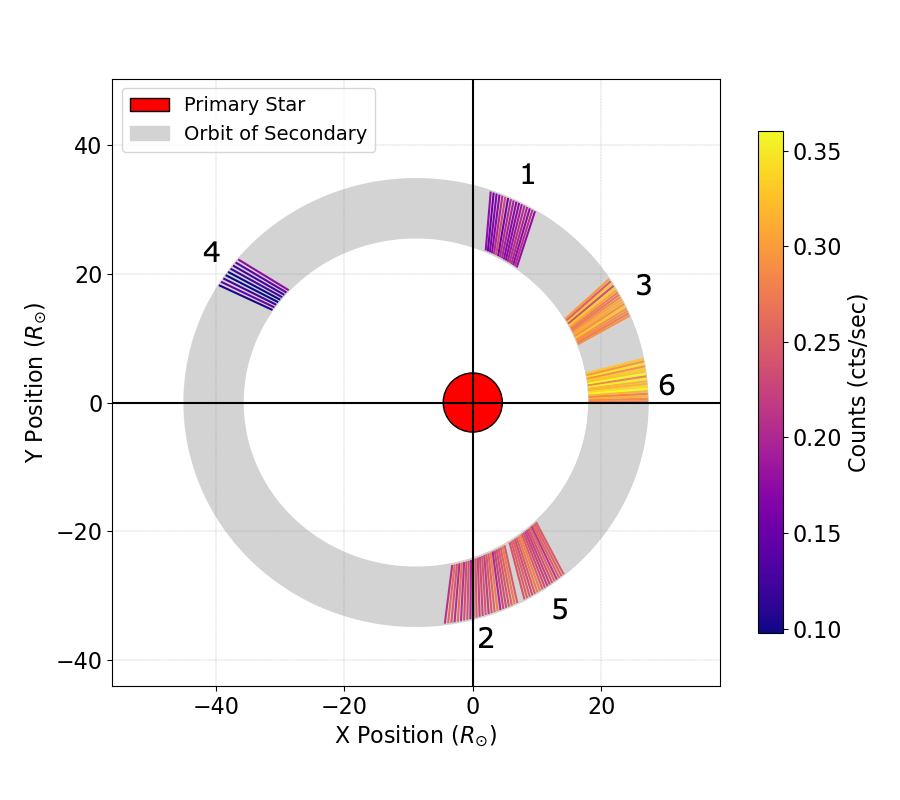}
    \caption{Representation of the orbit of $\epsilon$ Lupi (to scale). The red circle represents the primary, with an approximate radius of $4.6 \ R_{\odot}$. The grey track represents the orbit of the secondary, with width representing the secondary's radius, highlighting the proximity of the components with respect to their size. The distances are in solar radii. The coloured bands represent the position of our XMM-Newton observations reported in this paper. The color scale shows the count rate at each phase, binned every 15 minutes. The observation IDs from Table \ref{tab:XMM_obs1} are indicated around the circle.}
    \label{fig:lupi_orbit}
\end{figure}

In colliding wind binary (CWB) systems, the winds usually reach terminal velocities. In turn, this higher pre-shock velocity can produce even harder X-rays (up to $\sim$10 keV) (see \citealt{Rauw2016} for a review). Although the majority of known OB binaries display similar $L_{\rm x}/L_{\rm bol}$ ratios to those of single O-stars, some binary systems show a slightly higher $L_{\rm x}/L_{\rm bol}$ ratio, especially at higher energy bands \citep{Antokhin2008, Naze2009, Naze2011, Rauw2014}. The radiation field and gravity of both stars affect the shocks in colliding wind binaries \citep{Parkin2014}. In a magnetohydrodynamic simulation, \cite{Falceta-Goncalves2012} showed that the magnetic field generated by turbulence in the wind collision zone reduces the strength of the shock from wind-wind collision. In some cases, the excess emission may not come from wind-wind collision at all and may originate due to the presence of magnetically confined plasma \citep{Naze2010}. In the O-star binary Plaskett's star (HD 47129), where the secondary has a strong magnetic field, it is predicted that the observed X-ray emission is a combined effect of shocks within the secondary's wind confined within its magnetosphere, as well as from the collision between the secondary's magnetosphere and the primary's wind \citep{Linder2006, Linder2008, Grunhut2013, Rauw2016}.

\begin{deluxetable*}{ccccccl}
\tablenum{1}
\tablecaption{Observation details of the XMM-Newton observations of $\epsilon$ Lupi. \label{tab:XMM_obs1}} 
\tablewidth{0pt}
\tablehead{
\colhead{Data ID} &  \colhead{Obs ID} & \colhead{Revolution}  &  \colhead{Start UTC} & \colhead{Orb. Phase}  &  \colhead{Durations} & \colhead{Remarks} 
}
\startdata
0 & 0690210201 & 2424 & 2013-03-04 23:00:22 & $0.094 \pm 0.007$ & 7919 & Archival Observation \\
1 & 0904410401 & 4155 & 2022-08-17 05:51:00 & $0.218 \pm 0.018$ & 18000 &  \\
2 & 0904410501 & 4156 & 2022-08-19 15:51:42 & $0.757 \pm 0.027$*  & 38850 & *Only usable phase range mentioned \\
3 & 0904410601 & 4157 & 2022-08-21 06:13:04 & $0.098 \pm 0.018$ & 18000 &  Follow-up of archival obs. \\
4 & 0904410701 & 4158 & 2022-08-22 18:35:38 & $0.422 \pm 0.009$ & 11000 & Near-apastron observation \\
5 & 0904410801 & 4161 & 2022-08-29 01:25:06 & $0.814 \pm 0.023$ & 22000 & \\
\textbf{6} & \textbf{0904410901} & 4166 & 2022-09-08 00:18:31 & $\bm{0.007 \pm 0.033}$ & 30000 & \textbf{Periastron observation} \\ 
\enddata
\end{deluxetable*}

\subsection[epsilon Lupi]{$\epsilon$ Lupi}

$\epsilon$ Lupi A (HD 136504) is a double-lined spectroscopic binary consisting of two B-type stars in an eccentric orbit (e$\sim$0.28) with a period of $\sim4.6$ days \citep{moore1911, campbell1928, Thackeray1970, Uytterhoeven2005}. Figure \ref{fig:lupi_orbit} represents an approximate view of the binary orbit highlighting the proximity of the components. The component stars have similar masses ($M_{\rm pri} = 11.0_{-2.2}^{2.9} \ M_{\odot}$, $M_{\rm sec} = 9.2_{-1.9}^{2.4} \ M_{\odot}$), radii ($R_{\rm pri} = 4.64_{-0.48}^{0.37} \ R_{\odot}$, $R_{\rm sec} = 4.83_{-0.46}^{0.42}  \ R_{\odot}$), and temperatures ($T_{\rm pri} = 3407_{-567}^{658}$ kK, $T_{\rm sec} = 2197_{-399}^{489}$ kK, \citealt{Pablo2019}). This binary is part of a triple system with the tertiary component, $\epsilon$ Lupi B, orbiting with a period of $\approx740$ yr \citep{Zirm2007, Pablo2019}, and a separation of $\approx0.2$ arcsec.  Due to its large physical separation, the significance of this tertiary component in the context of our study is irrelevant, and henceforth, we refer to $\epsilon$ Lupi A as simply $\epsilon$ Lupi.

In the BinaMIcS survey, magnetic fields have been found in only one of the two components in all massive binary systems, except for the unique case of $\epsilon$ Lupi \citep{Shultz2015}. \cite{Shultz2015} derived dipolar surface magnetic field strengths ($B_{\rm d}$) of approximately 0.8 kG and 0.5 kG for the primary and secondary components, respectively. To date, only two other early-type binaries are known to host two magnetic components: HD 156424 \citep{Shultz2021} and BD+40 175 \citep{Elkin1999, Semenko2011}. However, $\epsilon$ Lupi A is the only known magnetic close binary with the potential for significant tidal and magnetospheric interactions, as the other two doubly-magnetic systems have much longer orbital periods (years to decades).

\cite{Pablo2019} found a heart-beat type light curve from optical observations, often observed in systems with high eccentricities, caused by tidal distortion. While the orbital characteristics of $\epsilon$ Lupi have been well-documented through numerous radial velocity measurements, the rotational periods of the two components remain unknown \citep{Pablo2019}. Weak temporal modulation in the longitudinal magnetic fields $\langle B_{\rm z} \rangle$ of the components suggests that the magnetic axes are closely aligned with the rotational axes, resulting in nearly-constant positive and negative $\langle B_{\rm z} \rangle$ values for the secondary and primary components, respectively \citep{Pablo2019, Biswas2023}. This anti-alignment is indeed expected from magneto-tidal interaction for this system \citep{Pablo2019}.

\begin{figure}
    \centering
    \includegraphics[width=0.4\textwidth]{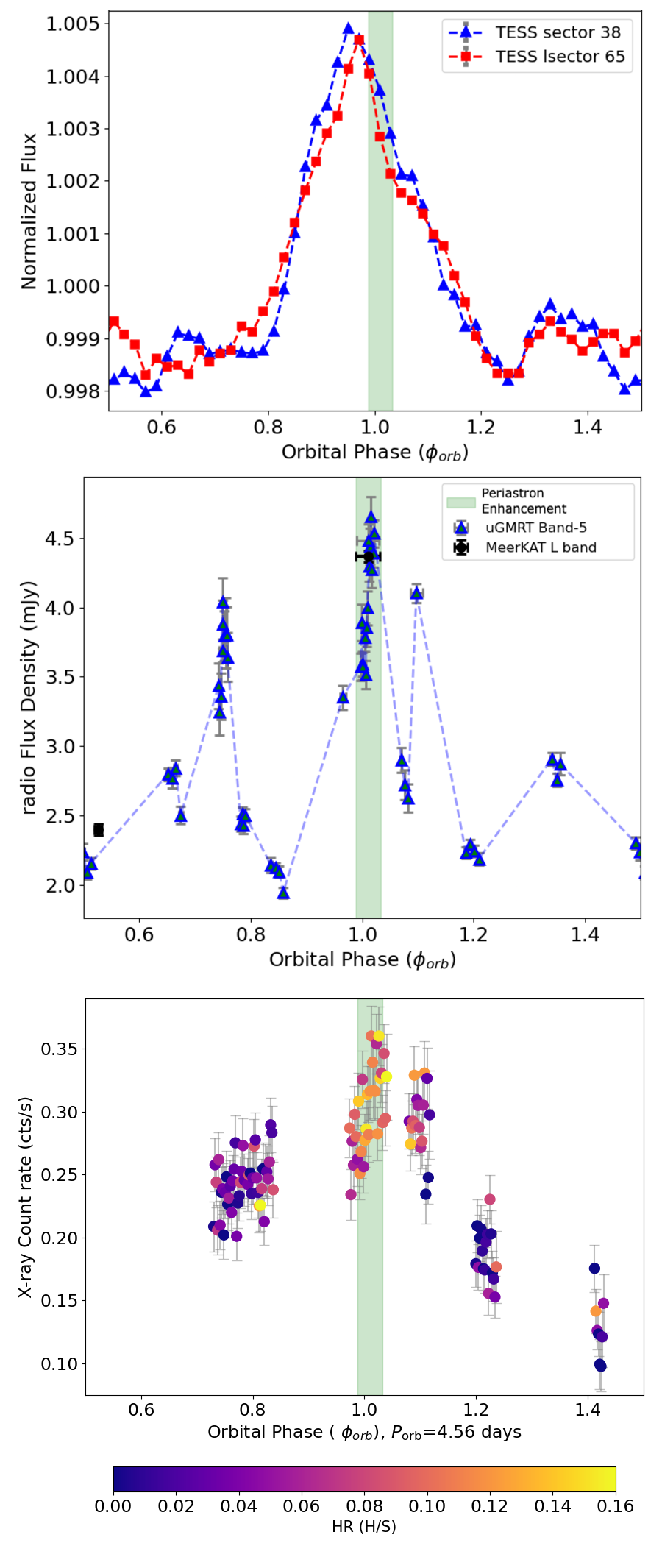}
    \caption{Orbital phase-folded light curve of $\epsilon$ Lupi in optical (top), radio (middle) and X-rays (bottom). The optical light curve is obtained by phase averaging the TESS light curve every $\Delta \phi_{\rm orb} = 0.01$ phase. The radio light curve in the middle is reproduced from \cite{Biswas2023} showing the radio light curve obtained with uGMRT and MeerKAT in the frequency range 1.15-1.45 GHz. The bottom plot shows the X-ray  count rate light curve obtained from EPIC-pn instrument. The color scale for the bottom figure represents the Hardness Ratio (HR). HR is defined as (count in 2.0-10.0 keV band)/(count in 0.2-2.0 keV band).  }
    \label{fig:orbital_compare}
\end{figure}

\cite{Biswas2023} detected variable radio emission from $\epsilon$ Lupi, with the radio light curve showing the existence of sharp radio enhancements at four orbital phases: one during the periastron phase and three others at orbital phases 0.09, 0.61, and 0.75 in bands $>1$ GHz (Fig. \ref{fig:orbital_compare}; middle panel). The enhancement at periastron was found to be persistent, the strongest, and the only enhancement in which linear polarization was detected. This behavior of enhanced flux during periastron was primarily attributed to synchrotron emission produced as a result of magnetospheric interaction between the two components via magnetic reconnection due to proximity.

This system was successfully detected in X-rays, first using XMM-Newton \citep{Naze2014}, and later using the Neutron Star Interior Composition Explorer mission (NICER) X-ray Telescope \citep{Das2023}. \cite{Das2023} detected variable X-ray emission from $\epsilon$ Lupi with NICER. Similar to the radio observations, the X-ray light curve also showed elevated flux near periastron compared to apastron. They characterized the X-ray enhancement as being confined to a very narrow phase range of $\sim$0.05 cycles. They concluded that the most advantageous situation involves magnetic reconnection initiated by the relative motion of the magnetospheres during specific orbital configurations. \cite{Biswas2023} also reported the existence of a very short-duration X-ray enhancement in the archival XMM-Newton X-ray light curve.

$\epsilon$ Lupi thus stands as the most prominent candidate for studying the combined effect of binarity and magnetic fields, particularly due to its stable fields and minimum-energy orbital configuration \citep{Pablo2019}. In this work, we focus on further studying X-ray emission from this intriguing system, covering nearly full orbital phases using XMM-Newton, which has much higher sensitivity and spatial/spectral resolution. 

This paper is structured as follows: the observation details and brief data analysis procedures are explained in Section \ref{observation}. In Section \ref{results}, we present results from the optical and X-ray observations, along with further details on X-ray data analysis. In Section \ref{discussion}, we discuss the findings from the X-ray results, and finally conclude in Section \ref{conclusion}.

\section{Observations} \label{observation}

The archival XMM-Newton observation of $\epsilon$ Lupi reported by \cite{Naze2014} was taken on 2013 March 4 (ObsID: 0690210201, PI: Y. Nazé). It spanned a brief phase range of $\sim 0.086-0.101$ ($\Delta \phi_{\rm orb} \approx 0.015$). We followed up this target with XMM-Newton during cycle AO-21 with a total effective observing time of $\sim$ 120 ks. The observations took place between 17 Aug 2022 and 9 Sep 2022, during revolution 4155 to 4166 (Prop. ID: 090441; PI: A. Biswas). The phase coverage of our observations can be visualized in Fig. \ref{fig:lupi_orbit}, and details of the observations are tabulated in Table \ref{tab:XMM_obs1}. The EPIC instrument\footnote{\url{https://www.cosmos.esa.int/web/xmm-newton/technical-details-epic}} was used in full-frame mode as the primary instrument, accompanied by the RGS instrument\footnote{\url{https://www.cosmos.esa.int/web/xmm-newton/technical-details-rgs}} in default spectroscopy mode. The thick filter was used while operating. As the target was optically bright, the optical monitor (OM) was turned off.

We carried out the data reduction using the XMM–Newton Science Analysis System (SAS, Version: 21.0.0)\footnote{\url{https://www.cosmos.esa.int/web/xmm-newton/sas}}. The tasks `{\it emproc}', `{\it epproc}', and `{\it rgsproc}' were used respectively to process EPIC-MOS, EPIC-PN, and RGS data. Data were filtered by selecting events only with a pattern below 12 (for EPIC-MOS) or below 4 (for EPIC-PN). Based on the light curve for events with energies between 10 and 12 keV and a time bin of 10 seconds, Good Time Intervals (GTI) were calculated with the selection criteria: PN rate $\leq$ 0.4 cts/sec and MOS rate $\leq$ 0.35 cts/sec. The background flaring was minimal. The possibility of pile-up was assessed using the task ‘{\it epatplot}', but significant pile-up was not predicted. During the second observation (data ID 2; revolution 4156), the detectors were kept on beyond the scheduled observing window for a few ks. However, upon examination, we found that these extra data were not usable and were thus excluded from further analysis.

We investigated light curves from the Transiting Exoplanet Survey Satellite (TESS) space telescope \citep{Ricker2014}. Each year, TESS observes 13 sectors for 27 days each, with a nominal precision of 60 ppm h$^{-1}$ in the wavelength range 600-1050 nm. Our target was observed during sector 38 (28 Apr 2021 to 26 May 2021) and sector 65 (04 May 2023 to 02 Jun 2023) with a 2-minute cadence. The processed 2-minute cadence light curves were obtained from the Mikulski Archive for Space Telescopes (MAST, \url{https://archive.stsci.edu/}). The light curve from sector 38 was previously reported by \cite{Das2023}. The XMM-Newton data was obtained in between sector 38 and 65. Thus we were able to confirm the orbital phasing with the help of these observations.

As this binary system exhibits apsidal motion \citep{Pablo2019}, we corrected the observed phases accordingly, given the orbital period ($P_{\rm orb}=4.559646^{0.000005}_{-0.000008}$), rate of periastron advance ($\Dot{\omega} = 1.1 \pm 0.1$), and reference periastron phase (${\rm T}_0 = 2439379.875^{0.024}_{-0.019}$). The initial corrected phase ($\phi_{\rm orb, 0}$) can be calculated using:
\begin{equation}
    \phi_{\rm orb, 0} = \left[ \left( \frac{{\rm HJD} - {\rm T}_0}{P_{\rm orb}} \right) \bmod 1 \right] - \left[ \left( \frac{{\rm HJD} - {\rm T}_0}{365.25} \cdot \frac{\Dot{\omega}}{360} \right) \bmod 1 \right],
\end{equation}
and finally the final corrected phase:
$$
\phi_{\rm orb}=\begin{cases}
			\phi_{\rm orb, 0}, & \text{if $\phi_{\rm orb, 0} \geq 0$}\\
            \phi_{\rm orb, 0} +1 , & \text{otherwise}
		 \end{cases}
$$

\section{Data Reduction and Results} \label{results}

\subsection{X-ray Count Rate Light Curve}

To generate background-subtracted X-ray light curves, we manually extracted unique source and background regions from the event files of individual observations to avoid contamination from nearby sources. The light curves were finally obtained using the Python wrapper for {\it SAS}, namely {\it pxsas}\footnote{\url{https://github.com/ruizca/pxsas}}. The EPIC-pn count-rate light curve, binned every 15 minutes, is shown in Fig. \ref{fig:orbital_compare} (bottom), where the color bar represents the variation of hardness ratio (HR). Here, we define HR  as the ratio of counts in the hard band (2.0-10.0 keV) to the soft band (0.2-2.0 keV). The HR plot mimics the total count rate light curve, with maximum during periastron, implying a gradual increase in hard X-ray emission towards the periastron.  For comparison with optical observation, the phase-folded (and apsidal motion corrected) TESS light curve  is shown in Fig. \ref{fig:orbital_compare} (top panel). The optical modulation observed in this case is due to the heartbeat phenomenon \citep{Pablo2019}, not an effect of the magnetosphere. The slight peak offset is a possible effect of the uncertainty in the rate of periastron advance, and subject to future investigation.

\begin{figure}
    \centering
    \includegraphics[width=0.45\textwidth]{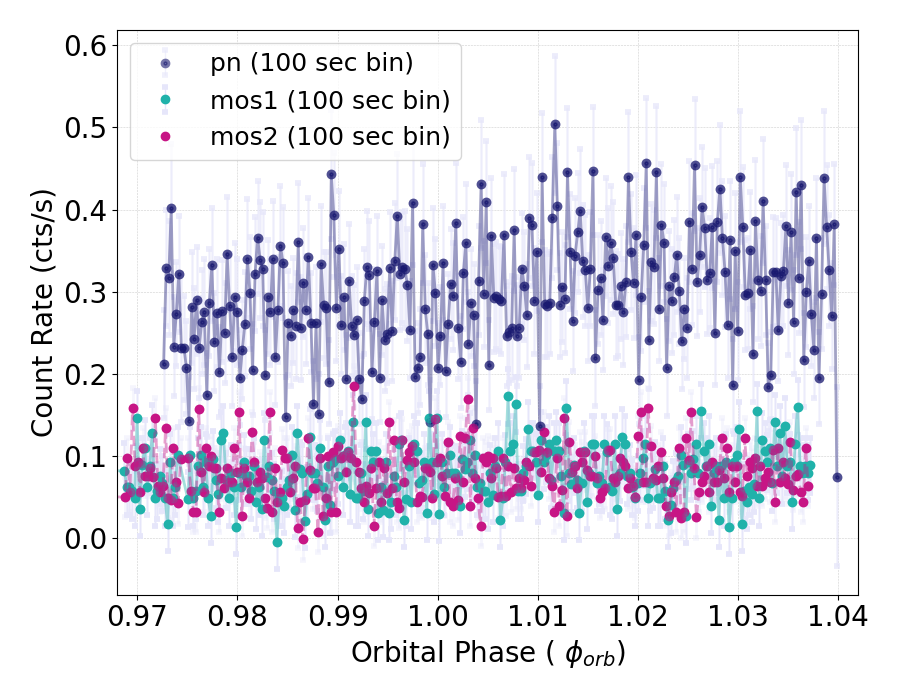}
    \caption{100-second binned EPIC light curve for the periastron observation (data ID:6). Neither the EPIC-pn, nor the EPIC-mos light curve show any short-duration enhancements.}
    \label{fig:peri_EPIC}
\end{figure}

The orbital phase-folded light curve shows a smooth variation, with the highest flux near periastron, exhibiting similar behavior to the optical light curves. A period analysis of the 15-minute binned light curve identified the orbital frequency as the only significant frequency. We did not observe any short-term enhancements ($\Delta\phi_{\rm orb}<0.01$) in any observation similar to the X-ray enhancement observed in the archival data reported by \cite{Biswas2023}, despite having similar instrument setup and analysis procedure. The periastron observation also showed a smooth increasing trend. EPIC-pn (referred to as `pn' henceforth) and EPIC-mos (`mos' henceforth) light curves with a short binning of 100 seconds are shown in Fig. \ref{fig:peri_EPIC}. Neither the light curve showed significant evidence of short-term enhancements compared to a flat model (p-values $>0.1$).

\subsection{EPIC Spectra}


We followed several approaches to model the EPIC spectra. Due to their high signal-to-noise, we primarily considered the pn spectra. The spectral analysis were performed with the \textit{xspec\footnote{\url{https://heasarc.gsfc.nasa.gov/xanadu/xspec/}}} package v.12.13.1 \citep{Arnaud1996}, along with its python wrapper \textit{pyxspec\footnote{\url{https://heasarc.gsfc.nasa.gov/xanadu/xspec/python/html/index.html}}}. For all the following methods, we used two multiplicative absorption models: one fixed absorption model accounting for ISM absorption (\textit{tbabs}), and another taking into account any circumstellar (e.g. magnetospheric) absorption (\textit{phabs}). In the following sections, we discuss two primary approaches: an optically thin thermal plasma model (\textit{apec}) with multiple temperature components (4T model), and a second model comprising thermal and non-thermal components (2T+pow model). For each data set, a fixed count per spectral bin was maintained, making the total number of bins different among datasets. However, we used a fixed range of photon energies while fitting the spectra (0.2--10 keV).

\begin{figure*}
\centering
\gridline{\fig{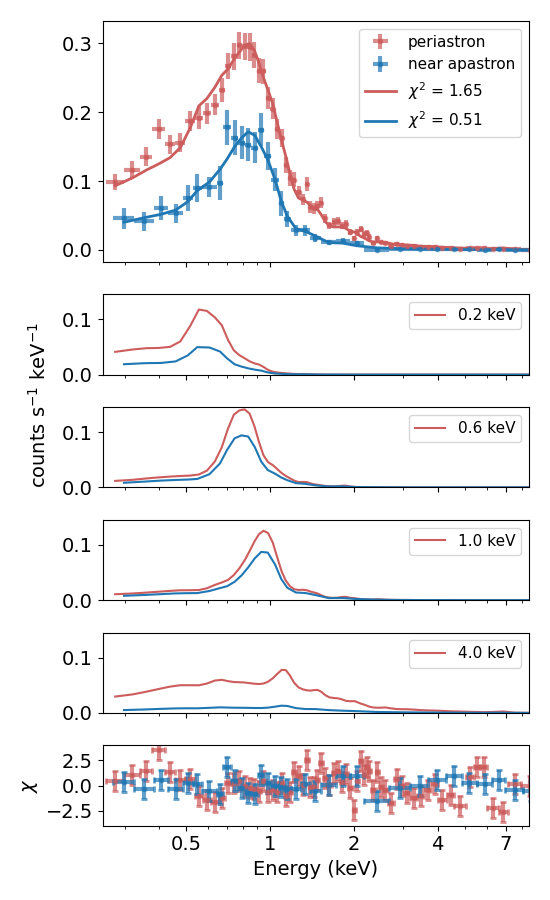}{0.45\textwidth}{(a) \textbf{4T} fit and model components} \label{fig:com1}
          \fig{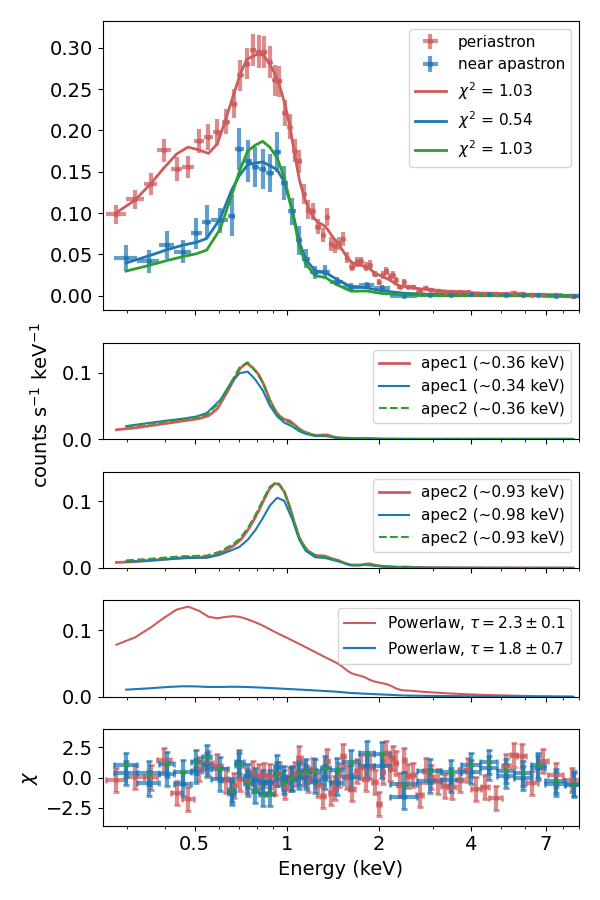}{0.45\textwidth}{(b) \textbf{2T+pow} fit and model components} \label{fig:com2}}
    \caption{EPIC-pn spectra of the periastron, and near-periastron observation with (a) 4T fit, and (b) 2T+pow fit. For both figures, the top panel shows the observed binned spectra, along with the corresponding model fit, as described in the text. For figure (a), middle-four panels show the comparison of model flux contribution from 0.2 keV, 0.6 keV, 1.0 keV, and 4.0 keV \textit{apec} models, respectively. For figure (b), the three middle panels show the model contributions from low-temperature \textit{apec}, high temperature \textit{apec}, and power-law component, respectively. We also over-plot a 2-T fitting for near-apastron phase in green, which gives a better fit. For both figures, the bottom panel shows $\chi$, which is defined as $\chi$ = (data-model)/residual.}
    \label{fig:spectra_compare}
\end{figure*}

\begin{deluxetable}{ccccccccccc}
\tablenum{2}
\tablecaption{The spectral fitting results from the 4-temperature (4T) fitting in \textit{xspec}. The $n_{\rm H}$ reported in this table correspond to the additional absorption model, rather than the fixed ISM absorption. The numbers in parenthesis in the $\chi^2_{\rm reduced}$ column represents the number of dof. The norm parameters are in units of $10^{-5}$ cm$^{-5}$. The used range of spectra for fitting is 0.2--10.0 keV. The flux represented here are the unabsorbed flux after ISM correction. \label{tab:XMM_fit}} 
\tablehead{
\colhead{Data} & \colhead{Mean} &  \colhead{$n_{\rm H}$} & \colhead{$\chi^2_{\rm reduced}$}  &  \multicolumn{4}{c}{Norm, N ($\times 10^{-5}$ cm$^{-5}$)}  &  \multicolumn{3}{c}{$\log (F_{\rm obs})$ (erg cm$^{-2}$ s$^{-1}$)}  \\
\colhead{ID} & \colhead{Phase} & \colhead{$\times 10^{22}$ cm$^{-2}$} & \colhead{} & \colhead{$N_{0.2 \ \rm keV}$} & \colhead{$N_{0.6 \ \rm keV}$} & \colhead{$N_{1.0 \ \rm keV}$} & \colhead{$N_{4.0 \ \rm keV}$} & \colhead{$\log F_{(\rm 0.2-10 \ keV)}$} & \colhead{$\log F_{(\rm 0.2-2 \  keV)}$} & \colhead{$\log F_{(\rm 2-10 \ keV)}$}
}
\rotate
\startdata
1 & 0.218 $\pm$ 0.018 & 0.00 $\pm$ 0.02 & 1.19(31) & 3.6 $\pm$ 1.1 & 4.4 $\pm$ 0.6 & 3.9 $\pm$ 0.5 & 2.5 $\pm$ 0.7 & -12.396 $\pm$ 0.020 & -12.437 $\pm$ 0.020 & -13.453 $\pm$ 0.020   \\
2 & 0.757 $\pm$ 0.027 & 0.00 $\pm$ 0.01 & 1.29(106) & 4.9 $\pm$ 0.9 & 4.0 $\pm$ 0.4 & 5.9 $\pm$ 0.4 & 7.8 $\pm$ 0.8 & -12.235 $\pm$ 0.006 & -12.309 $\pm$ 0.006 & -13.042 $\pm$ 0.012 \\
2a & 0.739 $\pm$ 0.009 & 0.00 $\pm$ 0.04 & 1.48(18) & 4.0 $\pm$ 2.9 & 4.2 $\pm$ 1.1 & 4.1 $\pm$ 1.0 & 9.1 $\pm$ 2.0 & -12.257 $\pm$ 0.028 & -12.351 $\pm$ 0.032 & -12.983 $\pm$ 0.020 \\
2b & 0.757 $\pm$ 0.009 & 0.00 $\pm$ 0.03 & 1.48(23) & 6.0 $\pm$ 2.5 & 3.9 $\pm$ 0.8 & 5.4 $\pm$ 1.0 & 6.1 $\pm$ 1.5 & -12.253 $\pm$ 0.027 & -12.313 $\pm$ 0.013 & -13.136 $\pm$ 0.026 \\
2c & 0.775 $\pm$ 0.009 & 0.00 $\pm$ 0.02 & 1.01(22) & 5.1 $\pm$ 1.4 & 4.0 $\pm$ 0.6 & 6.3 $\pm$ 1.0 & 6.8 $\pm$ 1.4 & -12.241 $\pm$ 0.015 & -12.305 $\pm$ 0.013 & -13.089 $\pm$ 0.014 \\
3 & 0.098 $\pm$ 0.018 & 0.00 $\pm$ 0.02 & 2.86(48) & 6.4 $\pm$ 1.4 & 2.1 $\pm$ 0.4 & 6.9 $\pm$ 0.6 & 16.1 $\pm$ 1.1 & -12.145 $\pm$ 0.015 & -12.257 $\pm$ 0.014 & -12.756 $\pm$ 0.014 \\
4 & 0.422 $\pm$ 0.009 & 0.00 $\pm$ 0.05 & 0.51(29) & 2.6 $\pm$ 1.7 & 2.3 $\pm$ 0.7 & 3.1 $\pm$ 0.7 & 3.8 $\pm$ 1.4 & -12.515 $\pm$ 0.037 & -12.583 $\pm$ 0.037 & -13.355 $\pm$ 0.037  \\
5 & 0.814 $\pm$ 0.023 & 0.00 $\pm$ 0.02 & 1.66(44) & 4.1 $\pm$ 1.2 & 4.4 $\pm$ 0.5 & 6.9 $\pm$ 0.6 & 4.4 $\pm$ 0.9 & -12.267 $\pm$ 0.016 & -12.319 $\pm$ 0.009 & -13.227 $\pm$ 0.016   \\
6 & 0.007 $\pm$ 0.033  & 0.00 $\pm$ 0.00 & 1.67(63) & 5.8 $\pm$ 0.4 & 3.4 $\pm$ 0.3 & 4.4 $\pm$ 0.4 & 21.7 $\pm$ 0.8 & -12.108 $\pm$ 0.010 & -12.238 $\pm$ 0.006 & -12.651 $\pm$ 0.009  \\
6a & 0.973 $\pm$ 0.011 & 0.00 $\pm$ 0.02 & 1.41(35) & 6.0 $\pm$ 1.8 & 4.3 $\pm$ 0.6 & 3.3 $\pm$ 0.7 & 17.2 $\pm$ 1.3 & -12.148 $\pm$ 0.018 & -12.264 $\pm$ 0.013 & -12.755 $\pm$ 0.018 \\
6b & 0.995 $\pm$ 0.011 & 0.00 $\pm$ 0.02 & 1.38(39) & 5.3 $\pm$ 1.4 & 3.3 $\pm$ 0.6 & 3.5 $\pm$ 0.8 & 25.2 $\pm$ 1.5 & -12.076 $\pm$ 0.010 & -12.236 $\pm$ 0.010 & -12.602 $\pm$ 0.017 \\
6c & 1.017 $\pm$ 0.011 & 0.00 $\pm$ 0.00 & 1.43(39) & 5.7 $\pm$ 0.7 & 2.4 $\pm$ 0.6 & 5.1 $\pm$ 0.7 & 27.1 $\pm$ 1.6 & -12.054 $\pm$ 0.010 & -12.223 $\pm$ 0.016 & -12.561 $\pm$ 0.016 \\
\enddata
\end{deluxetable}

\begin{deluxetable}{cccccccccccc}
\tablenum{3}
\tablecaption{Same as Table \ref{tab:XMM_fit}, but with the spectral fitting results from the 2T+pow fitting in \textit{xspec}. The phase range for each data ID is the same as Table \ref{tab:XMM_fit}. The $n_{\rm H}$ reported in this table also correspond to the additional absorption model, rather than the fixed ISM absorption. Here, T$_{\rm apec}$ represents the best-fit temperatures from \textit{apec} models. $\alpha$ represents photon index obtained from power-law model. In this case also, the flux values represent the ISM corrected unabsorbed flux.  \label{tab:XMM_fit_pow}} 
\tablehead{
\colhead{Data} &  \colhead{$n_{\rm H}$} & \colhead{$\chi^2_{\rm red}$}  &  \multicolumn{2}{c}{T$_{\rm apec}$ (keV)} & \colhead{$\alpha$} & \multicolumn{3}{c}{Norm, N ($\times 10^{-5}$ cm$^{-5}$)}  &  \multicolumn{2}{c}{$\log (F_{\rm obs})$ (erg cm$^{-2}$ s$^{-1}$)}  \\
\colhead{ID}  &  \colhead{$\times 10^{22}$ cm$^{-2}$} & \colhead{} & \colhead{kT$_1$} & \colhead{kT$_2$} & \colhead{} & \colhead{$N_{\rm kT1}$} & \colhead{$N_{\rm kT1}$} & \colhead{$N_{\rm pow}$} & \colhead{$\log F_{(\rm 0.2-10 \ keV)}$} & \colhead{$\log F_{(\rm 0.2-2 \  keV)}$} & \colhead{$\log F_{(\rm 2-10 \ keV)}$}
}
\rotate
\startdata
1 & 0.07 $\pm$ 0.04 & 0.60(29) & 0.37 $\pm$ 0.06 & 0.84 $\pm$ 0.05 & 3.4 $\pm$ 0.4 & 5.4 $\pm$ 1.8 & 5.4 $\pm$ 1.0 & 3.9 $\pm$ 0.9 & -12.123 $\pm$ 0.020 & -12.137 $\pm$ 0.020 & -13.635 $\pm$ 0.020 \\
2 & 0.03 $\pm$ 0.02 & 1.15(104) & 0.40 $\pm$ 0.04 & 0.94 $\pm$ 0.03 & 2.8 $\pm$ 0.2 & 4.8 $\pm$ 0.8 & 6.6 $\pm$ 0.5 & 5.7 $\pm$ 0.8 & -12.130 $\pm$ 0.012 & -12.166 $\pm$ 0.012 & -13.224 $\pm$ 0.012 \\
2a & 0.06 $\pm$ 0.07 & 0.72(16) & 0.41 $\pm$ 0.12 & 0.92 $\pm$ 0.10 & 3.1 $\pm$ 0.7 & 5.3 $\pm$ 2.4 & 5.0 $\pm$ 1.2 & 6.9 $\pm$ 2.4 & -12.049 $\pm$ 0.029 & -12.073 $\pm$ 0.029 & -13.318 $\pm$ 0.029 \\
2b & 0.00 $\pm$ 0.00 & 1.05(21) & 0.36 $\pm$ 0.05  & 1.06 $\pm$ 0.06 &  2.7 $\pm$ 0.3 & 6.4 $\pm$ 0.9 & 6.4 $\pm$ 1.2 & 3.5 $\pm$ 0.8 & -12.234 $\pm$ 0.016 & -12.270 $\pm$ 0.025 & -13.335 $\pm$ 0.025 \\
2c & 0.00 $\pm$ 0.02 & 0.73(20) & 0.73 $\pm$ 0.06 & 1.37 $\pm$ 0.36 & 3.0 $\pm$ 0.4 & 7.0 $\pm$ 1.3 & 4.8 $\pm$ 2.3 & 3.8 $\pm$ 1.5 & -12.189 $\pm$ 0.021 & -12.222 $\pm$ 0.021 & -13.330 $\pm$ 0.021\\
3 & 0.02 $\pm$ 0.02 & 1.48(46) & 0.69 $\pm$ 0.05 & 1.53 $\pm$ 0.14 & 2.8 $\pm$ 0.2 & 4.7 $\pm$ 0.6 & 12.2 $\pm$ 2.6 & 7.2 $\pm$ 1.7 & -12.052 $\pm$ 0.013 & -12.111 $\pm$ 0.013 & -12.951 $\pm$ 0.013    \\
4 &  0.00 $\pm$ 0.06 & 0.51(27) & 0.33 $\pm$ 0.05 & 0.98 $\pm$ 0.08 & 1.9 $\pm$ 0.7 & 3.9 $\pm$ 1.7 & 3.6 $\pm$ 0.7 & 1.5 $\pm$ 0.9 & -12.507 $\pm$ 0.038 & -12.591 $\pm$ 0.037 & -13.267 $\pm$ 0.038  \\
5 & 0.05 $\pm$ 0.04 & 1.60(42) & 0.37 $\pm$ 0.03 & 0.98 $\pm$ 0.03 & 3.0 $\pm$ 0.4 & 7.5 $\pm$ 1.8 & 8.1 $\pm$ 0.7 & 3.6 $\pm$ 0.9 & -12.140 $\pm$ 0.016 & -12.165 $\pm$ 0.016 & -13.393 $\pm$ 0.016   \\
6 & 0.03 $\pm$ 0.01 & 1.03(61) & 0.36 $\pm$ 0.04 & 0.93 $\pm$ 0.04 & 2.3 $\pm$ 0.1 & 4.4 $\pm$ 0.9 & 4.4 $\pm$ 0.4 & 12.1 $\pm$ 1.0 & -12.006 $\pm$ 0.010 & -12.105 $\pm$ 0.010 & -12.695 $\pm$ 0.010  \\
6a & 0.06 $\pm$ 0.03 & 1.24(33) & 0.19 $\pm$ 0.04 & 0.70 $\pm$ 0.04 & 2.5 $\pm$ 0.2 & 5.8 $\pm$ 2.9 & 6.7 $\pm$ 0.9 & 11.1 $\pm$ 1.5 & -11.975 $\pm$ 0.018 & -12.041 $\pm$ 0.018 & -12.825 $\pm$ 0.018 \\
6b & 0.00 $\pm$ 0.02 & 1.23(37) & 0.58 $\pm$ 0.14 & 1.09 $\pm$ 0.17 & 2.1 $\pm$ 0.1 & 3.6 $\pm$ 0.9 & 3.4 $\pm$ 1.0 & 11.0 $\pm$ 1.5 & -12.054 $\pm$ 0.016 & -12.197 $\pm$ 0.016 & -12.619 $\pm$ 0.016 \\
6c & 0.05 $\pm$ 0.02 & 1.16(37) & 0.31 $\pm$ 0.05 & 1.03 $\pm$ 0.08 & 2.4 $\pm$ 0.1 & 5.3 $\pm$ 1.8 & 5.5 $\pm$ 0.9 & 14.7 $\pm$ 2.1 & -11.929 $\pm$ 0.016 & -12.025 $\pm$ 0.016 & -12.630 $\pm$ 0.016 \\
\enddata
\end{deluxetable}

\begin{figure*}[ht!]
\gridline{\fig{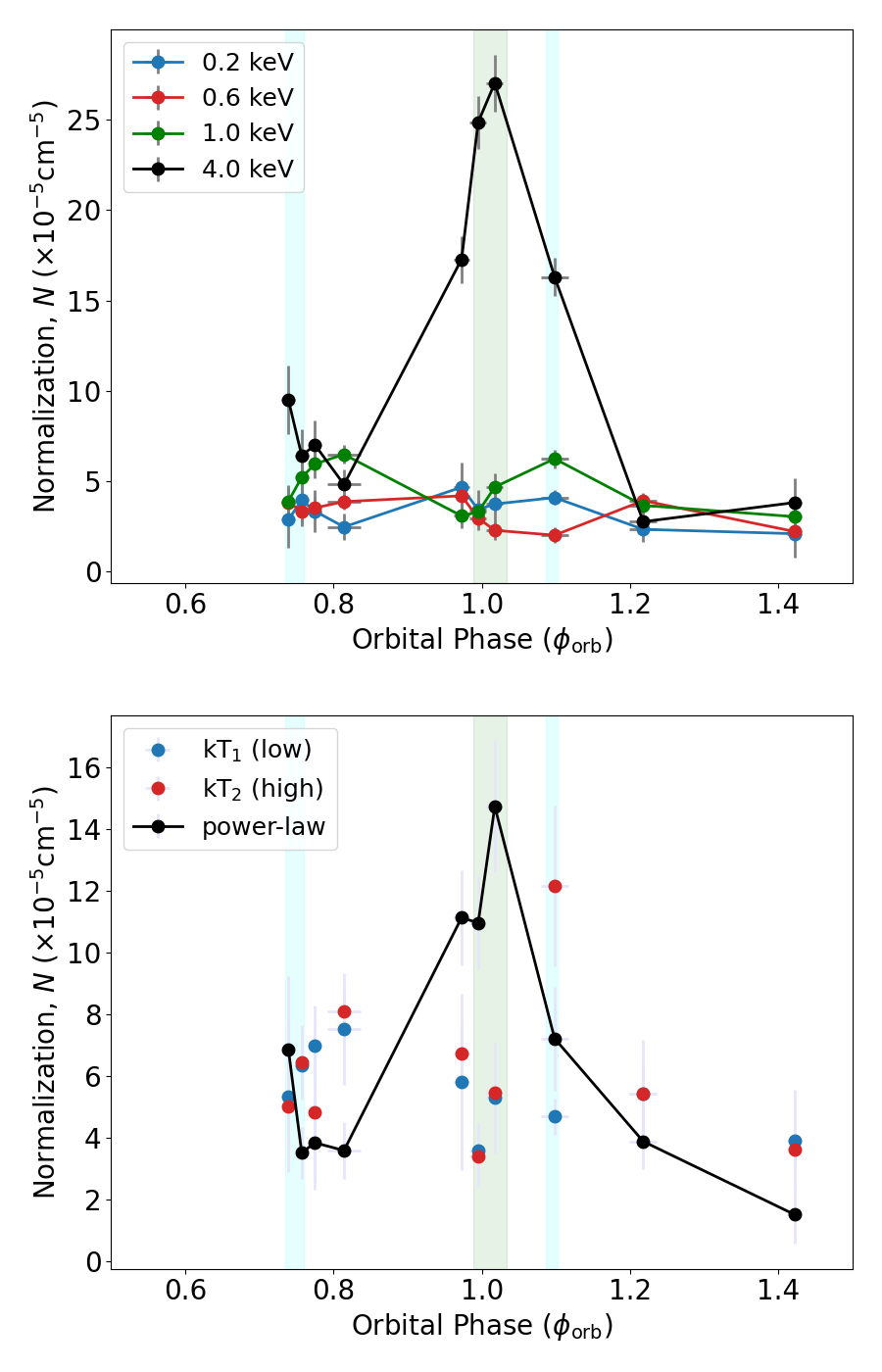}{0.44\textwidth}{(a) Orbital variation of model normalizations} \label{fig:norm_plot}
          \fig{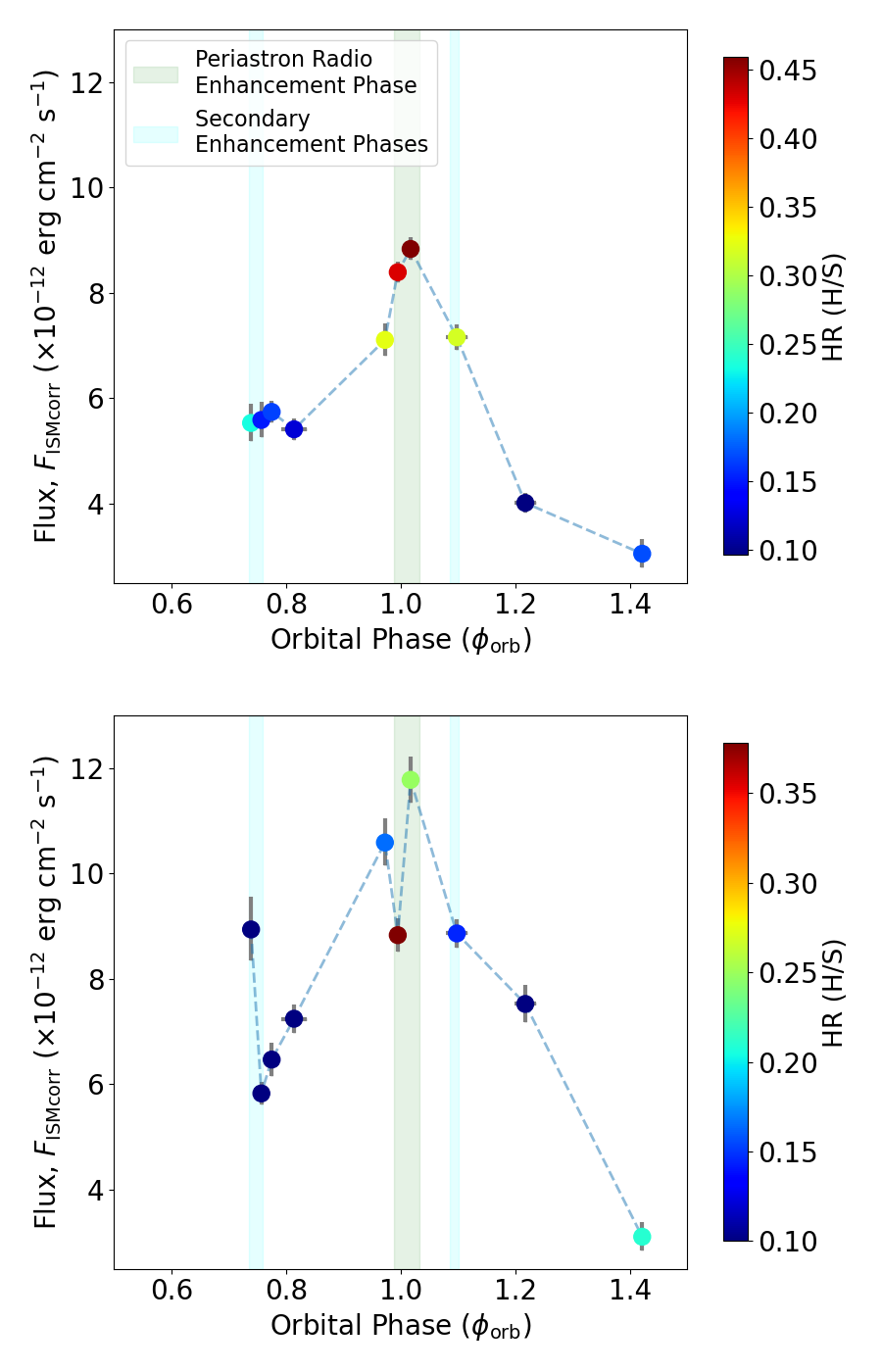}{0.44\textwidth}{(b) Orbital variation of fitted flux} \label{fig:flux_XMM}}
  \caption{Variation of normalization and X-ray flux with orbital period (for both models). (a) The variation of normalization parameters corresponding to the 0.2, 0.6, 1.0, and 4.0 keV temperatures from the 4T model (top), or the 2-temperature+powerlaw components from 2T+pow model (bottom)  with respect to orbital phase. (b) Orbital variation of fitted X-ray flux from the 4T (top), and \textit{2T+pow} model fitting (bottom). The green band near phase 1.0 (periastron) represent the phase of the main radio enhancement observed by \citealt{Biswas2023}. The cyan lines correspond to the phase range of two secondary enhancement observed in the radio light curve. \label{fig:orbital_variation_flux}}
\end{figure*}

\subsubsection{4T Model} \label{sec:4t}

 Firstly, we considered a set of four absorbed optically thin thermal plasma models (\textit{apec}), as utilized by \cite{Naze2014} and \cite{Das2023}. We refer to this method as the 4-temperature, or 4T model, with the four fixed temperatures being 0.2 keV, 0.6 keV, 1.0 keV, and 4.0 keV \footnote{The complete model expression used in \textit{xspec} is: `\textit{tbabs*phabs*(apec+apec+apec+apec)}'.}. These temperatures span the expected range of X-ray emission from magnetic massive binary systems. 

The interstellar medium (ISM) contribution for this system is known to be very small ($0.03 \times 10^{22}$ cm$^{-2}$, \citealt{Naze2014}). We fixed the ISM absorption component following \cite{Das2023}. The best-fit results from the 4T model are given in Table \ref{tab:XMM_fit}. We used the convolution model `cflux' to calculate the unabsorbed flux in the soft (0.2--2.0 keV), hard (2.0--10.0 keV), and full (0.2--10.0 keV) energy bands. The additional absorption model which was expected to take into account circumstellar absorption, was found to be consistently very small in all observations ($n_{\rm H}<10^{20}$ cm$^{-2}$), while their corresponding errors were larger (Table \ref{tab:XMM_fit}). Thus the absorbed and unabsorbed flux values were found to be of similar magnitude. In order to obtain a better understanding regarding the correlation and uncertainty of the fitted parameters, we performed a Markov Chain Monte Carlo (MCMC) analysis. Details about the MCMC analysis can be found in Appendix \ref{sec:mcmc}. We found that the 2D probability distributions of all parameters (except $n_{\rm H}$) are well constrained (see Fig. \ref{fig:A2}), and the associated uncertainties are similar to those we obtained with $\chi^2$ fitting in \textit{xspec}. However, it is worth mentioning that in some observations (ID 3, 5, and 6), high reduced $\chi^2$ were observed.

Figure \ref{fig:spectra_compare}(a) shows two sample spectra over-plotted in the topmost panel: one taken during periastron (red) and another taken near apastron (blue), along with the 4T fit to each dataset (solid lines of corresponding color). We compared the response of these two observations, and found them to be nearly identical. Thus, it is meaningful to directly compare these two spectra. The count rate is clearly much higher during periastron throughout the observed energy range, in all models. For better comparison, the y-range of all the panels showing model flux is kept fixed. The difference is most pronounced in the flux contribution from the 4.0 keV model, where the 4.0 keV \textit{apec} model during apastron has very little contribution to the overall spectra. This result was also expected from the hardness ratio illustrated in Fig. \ref{fig:orbital_compare}. 

This variation of harder X-rays is much more evident in the orbital variation of the relative contribution from the thermal plasma models, as shown in Fig. \ref{fig:orbital_variation_flux}a (top panel). This figure shows the orbital variation of the normalization parameters of each \textit{apec} model (0.2, 0.6, 1.0, and 4.0 keV). For data ID 2 and 6 (orbital phase $\sim0.76$ and $\sim0.0$), the observations being long enough, we split them in time into three separate spectra to further investigate the variation on a shorter timescale.  The normalization factors corresponding to low-temperature plasma ($\leq 1$ keV) remain similar across the orbital cycle. However, we noticed a jump in the normalization of the 4.0 keV component at periastron. During periastron observation, the normalization for 4.0 keV component was also found to be varying rapidly.

We show the orbital variation of fitted unabsorbed X-ray flux in the range 0.2--10.0 keV obtained from the 4T model in Fig. \ref{fig:orbital_variation_flux}b (top panel).  The flux variation exhibits a variability pattern similar to the count rate light curve, as expected. We observe a more than threefold increase in periastron flux compared to the fitted flux near apastron.

\subsubsection{`\textit{2T+pow}' Model} \label{sec:pow}

The detection of non-thermal radio emission with linear polarization during periastron by \cite{Biswas2023} motivates the search for non-thermal X-ray emission arising due to synchrotron and/or Inverse Compton mechanisms in our dataset. \cite{Ryspaeva2020} also proposed that for binary OB stars, a combination of thermal and power-law models may provide a superior fit. The power-law component surpasses the contribution of the thermal component for energy values $E>2$ keV. \cite{Leto2018} found comparable fit statistics with an only-\textit{apec} model and a combined \textit{apec+powerlaw} model for the non-degenerate strongly magnetic star HR 5907.

With these physical motivations, and with an attempt to improve the fit, we employed an alternative fitting strategy, integrating two thermal \textit{apec} models with an additional power-law component \footnote{The complete model expression used in this case is: `\textit{tbabs*phabs*(apec+apec+pow)}'.}. We refer to this as the 2T+pow model. In this case, we did not fix the temperatures but treated them as free parameters. We used absorption models similar to the 4T model. The fitted parameters and unabsorbed model flux values are tabulated in Table \ref{tab:XMM_fit_pow}.

The reduced $\chi^2$ improved significantly for the near-periastron observations, while they remained similar during near-apastron observations. For the periastron observation, the reduced $\chi^2$ improved from 1.63 in the 4T model fit to 1.03 in case of 2T+pow fit (Table \ref{tab:XMM_fit_pow}), formally indicating the latter to be an optimal fit. A similar trend is observed during the phase $\phi_{\rm orb} \sim 0.1$, where the 4T model exhibited a poor reduced $\chi^2$ of $\sim2.9$. However, the fit significantly improved with the 2T+pow model, yielding an improved reduced $\chi^2$ of $\sim1.5$. As these models have a different number of free parameters, to determine the statistically better-fitting model we performed an extra-sum-of-squares F-test. This test compares the residual sum of squares (RSS) of the models to evaluate if the more complex model significantly improves the fit over the simpler model. The F-statistic is calculated as follows:

\begin{equation}
F = \frac{(\text{RSS}_{4T} - \text{RSS}_{\text{pow}}) / (\text{df}_{4T} - \text{df}_{\text{pow}})}{\text{RSS}_{\text{pow}} / \text{df}_{\text{pow}}}
\end{equation}

\noindent where $\text{RSS}_{4T}$ and $\text{RSS}_{\text{pow}}$ are the residual sums of squares for the simpler (4T) and more complex (pow) models based on the total number of free parameters, respectively, $\text{df}_{4T} = N - V_{4T}$ and $\text{df}_{\text{pow}} = N - V_{\text{pow}}$ are the degrees of freedom for the models, $N$ is the number of data points, and $V_{4T}$ and $V_{\text{pow}}$ are the number of parameters in the simpler (4T) and more complex (pow) models, respectively. For our data, $V_{\text{4T}} = 5$, and $V_{\text{pow}} = 7$, we calculated the F-statistic and determined the p-value for each observation. We noticed F-values $>1$ for all observations, supporting the improvement of fit. For periastron observation, we obtain a p-value $\ll 0.01$, making the 2T+pow model an obviously better fit. This behaviour is not changed even if the temperatures in the 4T model are allowed to vary freely, which suggests that the power-law component is the primary reason for this better fit.

Similar to the 4T approach, comparisons of the 2T+pow fit to the periastron and apastron spectra along with their corresponding model contributions are shown in Fig. \ref{fig:spectra_compare}b. Although we allowed the temperatures for the two \textit{apec} models to be free parameters, they converged to the same values within errors. The normalization parameters for each of these thermal models are also similar, within their respective uncertainties. However, there is a striking difference in the contribution of the power-law component during periastron and apastron (see the 4th panel of Fig. \ref{fig:spectra_compare}b). While the power-law has a negligible contribution during apastron, it is the major contributor during the periastron phase. If we fit the near-apastron phase with only two `\textit{apec}' components, we obtain a better fit, with reduced $\chi^2 \sim 1$ (green curve in Fig. \ref{fig:spectra_compare}b). This treatment further demonstrates the irrelevance of power-law near apastron.

In the '2T+pow' model the contributions from the \textit{apec} components remain roughly constant during all observations (Fig. \ref{fig:orbital_variation_flux}a-bottom). The normalization of the power-law component increases rapidly from near-apastron to periastron phase. In Fig. \ref{fig:orbital_variation_flux}, the phase range for the primary radio enhancement at periastron (green band) and the secondary radio enhancements at two other phases (cyan bands) are highlighted. The fitted X-ray flux (with 2T+pow model) during these periods are also found to be elevated in the X-rays, similar to radio enhancements (see Fig. \ref{fig:orbital_variation_flux}b-bottom).

\begin{figure}
    \centering
    \includegraphics[width=0.48\textwidth]{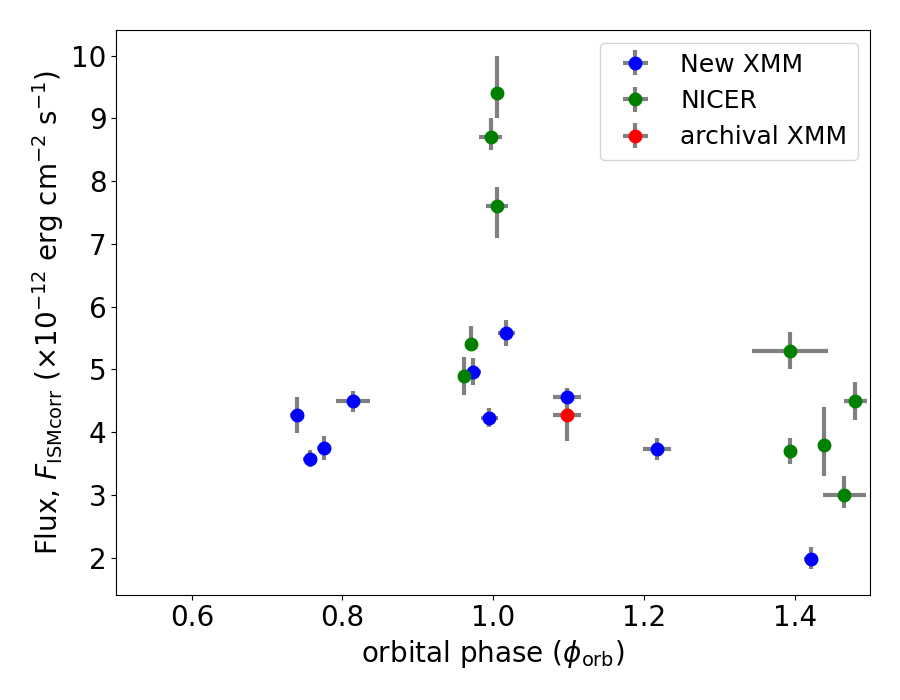}
    \caption{Plot comparing the fitted flux values from new observations (blue), archival observation (red) and NICER observations (green) in the energy range 0.5 -- 2.0 keV. NICER flux values were taken from \cite{Das2023}, and come from 4T fitting, while the XMM-Newton data are fitted with the 2T+pow model. The horizontal error bars represent the span of each observation.}
    \label{fig:NICER}
\end{figure}

\subsection{Comparison with NICER data}

Upon scrutinizing the X-ray spectra, \cite{Das2023} also identified striking differences between the X-ray emission during periastron  and near apastron. In both cases, when using the 4T \textit{apec} model, they observed a similar contribution in flux from low-energy thermal plasma ($\leq 1.0$ keV). However, a difference was noted in the relative (or overall) contribution of the 4.0 keV component, similar to the finding of this work. However, the high background in the NICER data and the poorer spatial/spectral resolution made it difficult for \cite{Das2023} to distinguish between different spectral models. The comparison of the orbital flux variations (with 4T model) from the NICER and XMM-Newton data is shown in Fig. \ref{fig:NICER}. The archival XMM-Newton data are also over-plotted in that figure.

While the pattern of the variability remains similar, the NICER fluxes are in general found to be much higher even in the same energy range (0.5 -- 2.0 keV, in this case). This discrepancy may arise due to the high background observed with NICER. In the XMM-Newton image, few nearby sources can be seen. As the spatial resolution is poor in case of NICER, these additional sources may also have contaminated the target spectra. Furthermore, it should be noted that the error bars of the NICER data (which are the 68\%, or 1-$\sigma$ confidence intervals calculated from MCMC analysis) are much larger than those of the XMM-Newton data.

\begin{figure}
    \centering
    \includegraphics[width=0.49\textwidth]{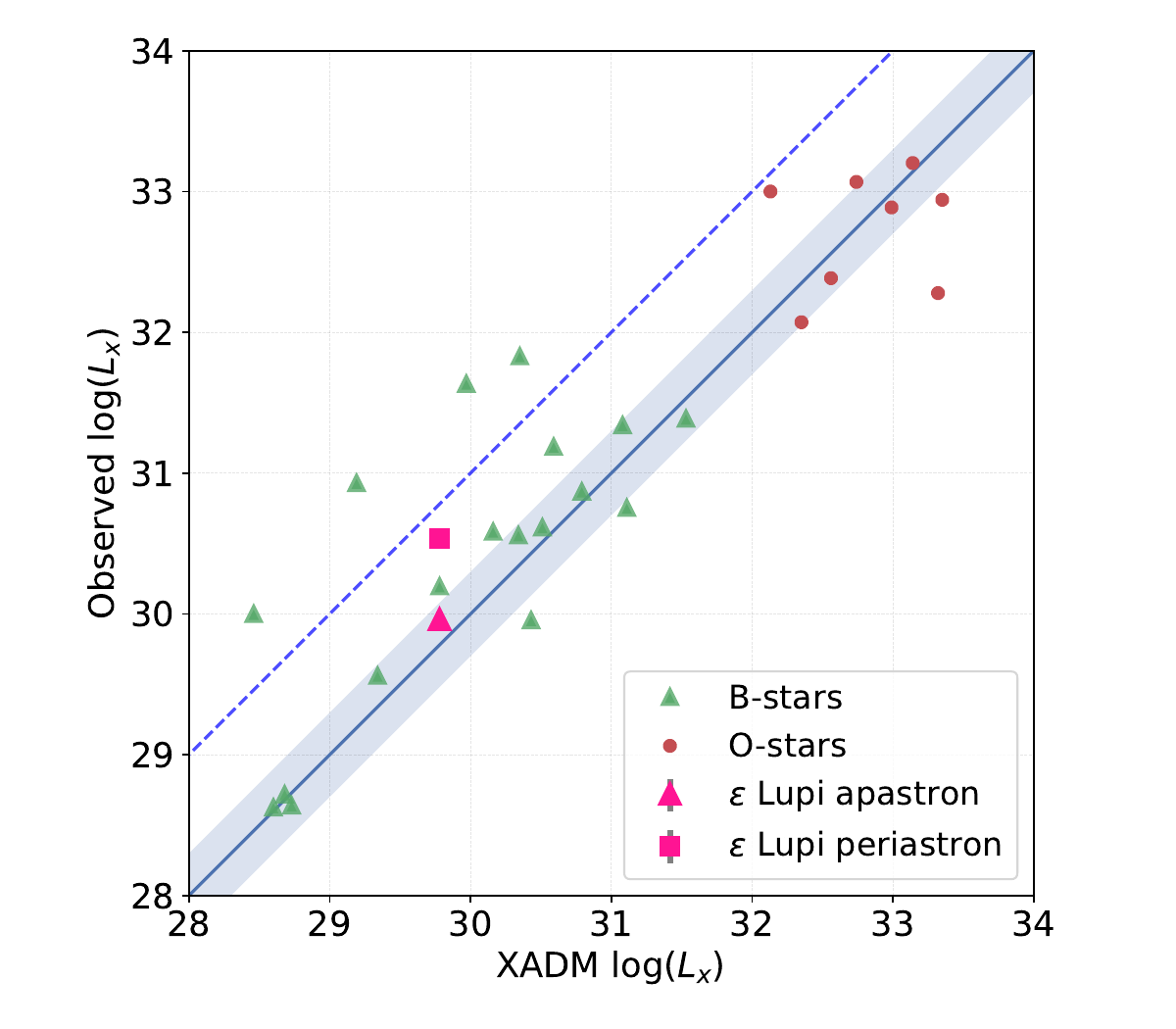}
    \caption{Position of near-apastron X-ray luminosity of $\epsilon$ Lupi in the comparison plot by \citealt{Naze2014} (see Fig 10 of \citealt{Naze2014})  between the observed X-ray luminosity of magnetic OB stars and the theoretical values using the XADM model of \citealt{ud-Doula2014}. The blue dotted line represent the ideal model with 100\% efficiency; the blue solid line correspond to a scaling by 10\%; and the light blue shaded area correspond to scaling by 5-20\% efficiency, which is consistent with magneto hydrodynamic (MHD) models. The pink triangle and rectangle represents the near-apastron, and periastron observation, respectively.}
    \label{fig:scaling}
\end{figure}

\section{Discussion} \label{discussion}

\subsection{Origin of Observed Emission: Single Star Emission vs Binary Interaction}

As previously discussed, magnetospheric interaction is anticipated in the case of $\epsilon$ Lupi, as the magnetospheres of the two B-type components are consistently in contact \citep{Shultz2015}, and the eccentricity of the orbit drives periodic stresses in the shared field topology. This phenomenon is manifested as enhanced radio emission, particularly during periastron \citep{Biswas2023}. The X-ray light curve (\citealt{Biswas2023, Das2023}, and this work), whether characterized by count rate or fitted flux, exhibits a similar trend: a variable nature with the highest flux occurring towards periastron. Motivated by the radio observations, the preliminary hypothesis is to attribute this variability to binarity interaction either due to wind-wind, or magnetospheric interaction.

In Sections \ref{sec:4t} and \ref{sec:pow}, we have seen that the contribution from softer thermal plasma remains similar throughout the orbit. The softer X-rays can be explained by X-ray emission originating in the individual stars' magnetospheres due to Magnetically-Confined Wind Shocks (MCWS), as discussed in Sec. \ref{sec:intro}. In fact, we investigated the position of $\epsilon$ Lupi in the context of the scaling relationship for X-ray emission from magnetic OB stars. \citealt{Naze2014} attempted to find the relation between observed X-ray luminosity and the theoretical value of the luminosity obtained from the 10\% efficiency `X-ray Analytical Dynamical Magnetosphere' (XADM) model developed by \cite{ud-Doula2014}. In Fig. \ref{fig:scaling}, we show the position of $\epsilon$ Lupi in the scaling relationship, with XADM luminosity calculated by \cite{Naze2014}. The pink triangle in Fig. \ref{fig:scaling} represents the  luminosity observed during the near-apastron observation. Stars well beyond the 10\% efficiency XADM line are characterized by the most extreme combinations of magnetic field and rotational velocity. However, $\epsilon$ Lupi is not an extreme case from either perspective. The X-ray luminosity near apastron is thus consistent with X-ray emission from two magnetospheres. However, the luminosity during periastron is significantly higher (pink rectangle in Fig. \ref{fig:scaling}), suggesting an additional origin, i.e., the role of binarity.

Hard X-ray emission resulting from binary interaction can arise from different mechanisms, namely colliding winds from two components or, in the case of doubly-magnetic systems, through magnetic reconnection. The winds from main sequence B-type stars are expected to be weak due to their low mass-loss rates, and there is also no strong observational evidence of excess X-ray emission from wind-wind interaction in main sequence B-type binaries \citep{Naze2011, Rauw2016}. Consequently, magnetospheric interaction becomes the more plausible option. \cite{Das2023} also suggested magnetic reconnection as a possible origin of the repeating X-ray enhancements near periastron.

Enhanced X-ray (or radio) emission during periastron has been reported for pre-main sequence (PMS) stars (e.g., \citealt{Massi2002, Massi2006, Gregory2014, Getman2016, Getman2023}). \cite{Getman2011} suggested that this additional X-ray emission can result from magnetic loops connecting two magnetic stars, akin to the model proposed by \cite{Uchida1985}, as well as from the magnetic loop of a single star. A magnetic loop connecting one magnetic body to another, creating a complex topology, may be responsible for additional energy in such doubly-magnetic systems \citep{Lanza2009}. PMS stars do not exhibit very stable enhancement properties, i.e. the periastron enhancements are not always regular, and the flux densities of these short enhancements also vary within cycles. These observed short-duration flares can be explained as a sudden release of magnetic stress. However, the radio enhancements of $\epsilon$ Lupi during periastron are very stable, with comparable fluxes observed during multiple periastron passages \citep{Biswas2023}, observed with different telescopes 1-2 years apart. This stability may be related to the fact that for $\epsilon$ Lupi, the Alfv{\'{e}}n radii ($R_{\rm A}$) are larger than the orbital separation, leading to a  variable supply of energy from the change in magnetic energy with orbital variation. As the stars approach periastron, magnetic reconnection intensifies, and the magnetic energy released varies with $d^{-3}$, where $d$ represents the orbital separation at any given instant. The power-law fit to the periastron data suggests a possible non-thermal component, which was negligible in the case of the near-apastron data. In this scenario, non-thermal emission would be generated at the site of magnetic reconnections occurring in the magnetic loops connecting the two stars.

\begin{figure}
    \centering
    \includegraphics[width=0.48\textwidth]{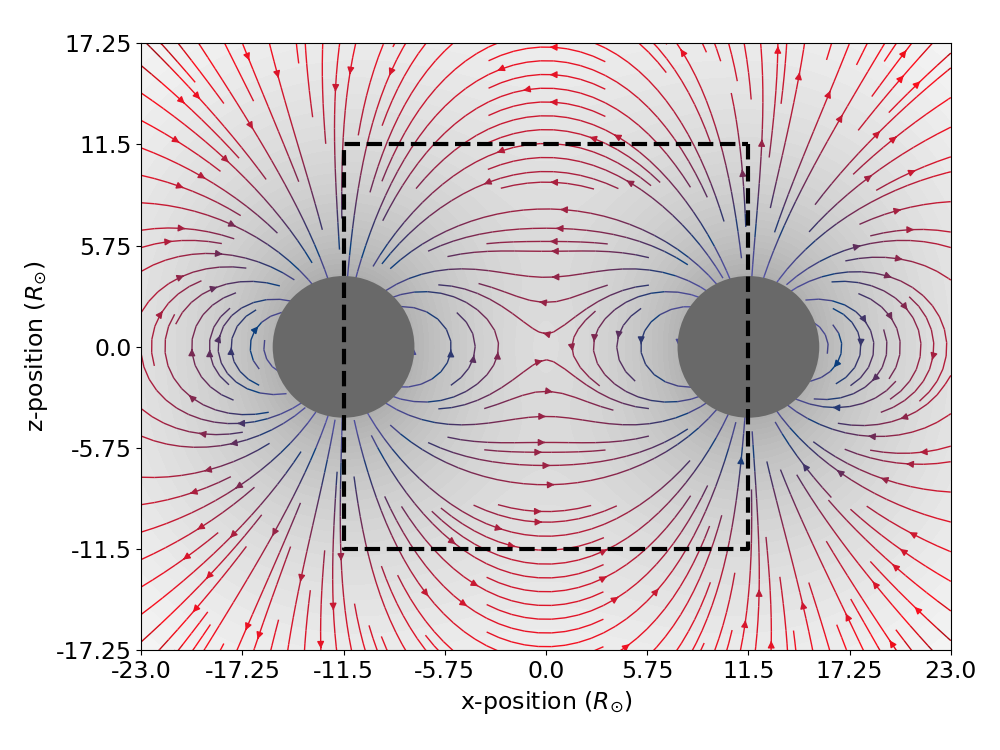}
    \caption{Composite magnetic field of the two components of $\epsilon$ Lupi, assuming a pure dipolar geometry. The black box represent the minimum region (2-D projection) of integration to calculate the interaction energy from equation \ref{eq:3}. The separation and the radius of the stars are to scale.}
    \label{fig:config_mag}
\end{figure}

\subsection{Magnetic Reconnection as the Most Feasible model}

To obtain a quantitative estimate of the magnetic energy available from reconnection at periastron, we follow the approach prescribed by \cite{Adams2011}.  We make the reasonable assumptions that the stars are perfectly anti-aligned \citep{Pablo2019}, ignoring rotation, i.e. any torque in the magnetospheres. If we approximate the stellar fields as pure dipoles with moments $\mathbf{m}_j=m_j \hat{z}$ ($i=1,2$), the magnetic fields of each component can be represented as:

\begin{equation}
    \mathbf{B}_j = \frac{m_j}{r_j^3} (2 \cos \theta \hat{r}_j + \sin \theta \hat{\theta}_j)
\end{equation}

The composite magnetic field can be visualized in Fig. \ref{fig:config_mag}. We can calculate the interaction energy, i.e. the difference in the stellar field energy as \citep{Adams2011}:

\begin{equation}
    E_{\rm int} = \int_V \frac{\mathbf{B}_1 \cdot \mathbf{B}_2}{4 \pi}  d^3 r - \int_{V_1} \frac{B_1^2}{8 \pi} d^3 r - \int_{V_2} \frac{B_2^2}{8 \pi} d^3 r,   \label{eq:3}
\end{equation}

\begin{figure}
    \centering
    \includegraphics[width=0.45\textwidth]{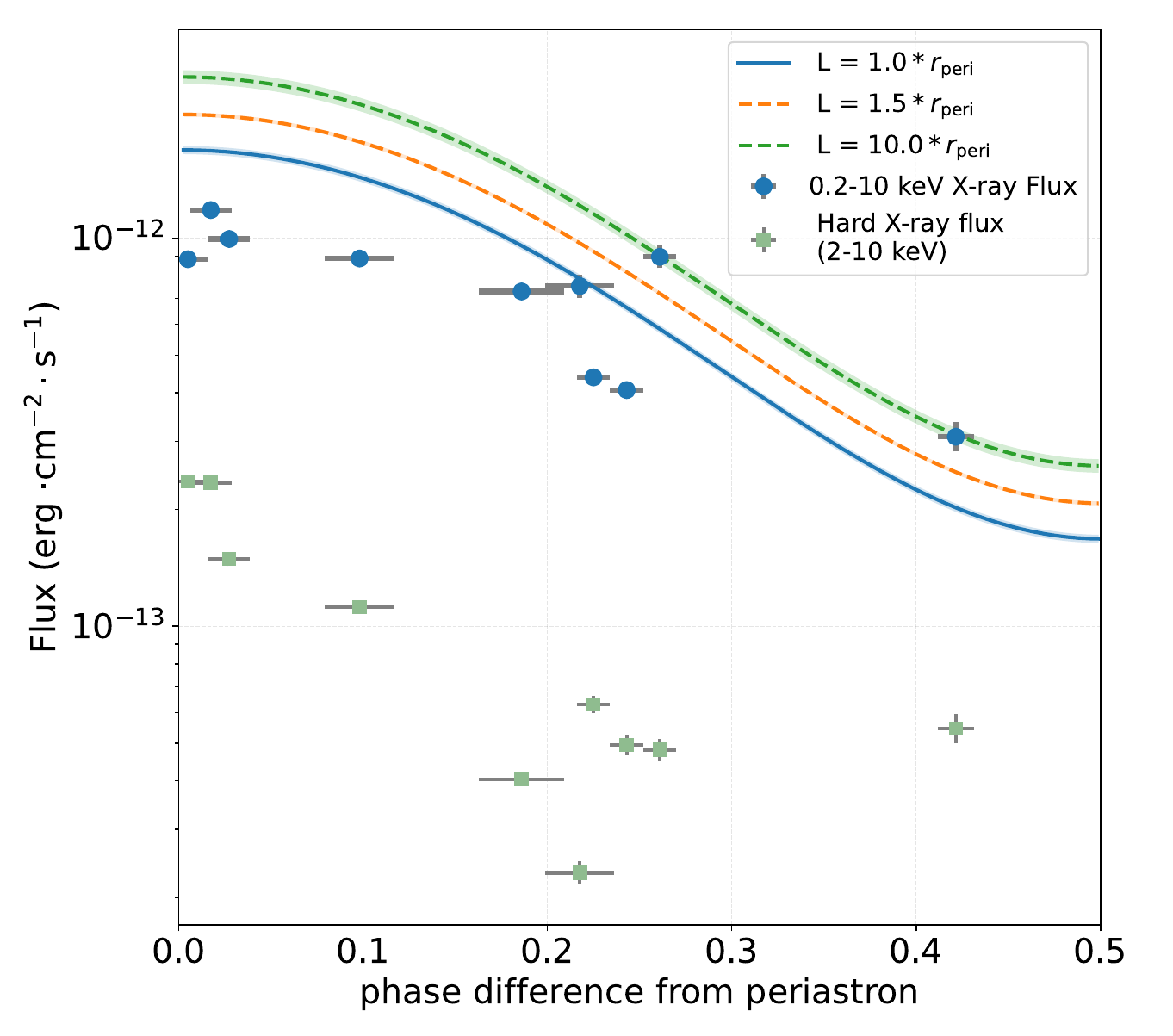}
    \caption{Variation of model flux obtained from magnetospheric reconnection within cubes of sides $L=$ 1, 1.5, and 10 times the periastron separation (curves), along with the observed X-ray flux in 0.2-10 keV band (blue circle), and hard X-ray in 2-10 keV band (green squares). 0 in the x-axis represent the periastron.}
    \label{fig:magnetic_model}
\end{figure}

\noindent
where $V_j$ represents the spherical volumes of the primary and secondary stars. We calculate the interaction energy and, consequently, the rate of change of energy using stellar parameters and magnetic field values from \cite{Shultz2015} and \cite{Pablo2019}. With the distance to $\epsilon$ Lupi obtained from the Hipparcos parallax \citep{van2007} at 156 parsecs \citep{Pablo2019}, we compute the flux available from magnetospheric interaction. The variation of model flux concerning the distance from periastron is depicted in Fig. \ref{fig:magnetic_model}. The different curves represent the amount of flux available by considering different effective regions of interaction.  It is evident that towards periastron, the magnetic field can supply enough power to fuel the excess hard X-ray emission. However, near apastron, the interaction energy is insufficient to fuel the X-ray emission. Thus during apastron, single star emission is the only major contributor towards the X-ray flux. This flux during apastron, as discussed before, can be explained with the XADM for solitary magnetic hot stars. \cite{Das2023} also gave a rough estimate of the power available in this system. However, the authors did not consider the exact geometry, and an unjust approximation of the closed magnetic loops being $\gg R_*$ by \cite{Adams2011} resulted in a large upper limit (two order of magnitude higher than this work).

\subsection{Additional Enhancements}

As the rotation periods of the components of $\epsilon$ Lupi are not known, a quantitative estimate of the contribution of rotational motion towards magnetospheric interaction is not feasible. Thus the model above does not account for the effect of rotation. Rotation can introduce twisting in flux tubes, leading to further reconnection events and producing more energy \citep{Uchida1985}.  Further enhancements can be caused by the twisting of these flux tubes, leading to a build-up of magnetic energy and eventual sudden release upon reaching an instability threshold \citep{Cherkis2021}. This phenomenon may explain the off-periastron enhancements observed in both radio \citep{Biswas2023} and X-rays (if one considers the 2T+pow model fitted flux light curve; Fig. \ref{fig:orbital_variation_flux}b).

\section{Conclusion}  \label{conclusion}

In this paper, we present a detailed analysis of XMM-Newton data obtained to investigate the X-ray emission from the unique doubly-magnetic hot binary $\epsilon$ Lupi. This investigation was prompted by the complementary study of radio emission presented by \cite{Biswas2023} and the X-ray study with NICER by \cite{Das2023}. Our XMM-Newton observations, spanning the orbital period of $\epsilon$ Lupi, revealed variable X-ray emission peaking near periastron. While the EPIC-pn light curve exhibited orbital variability, no distinct short-term enhancements were observed.

We applied a comprehensive spectral analysis, utilizing a 4-temperature thermal plasma model following \cite{Naze2014} and \cite{Das2023}. The analysis uncovered an excess of hard X-ray emission during periastron, which was also observed in the NICER data. We attributed the near-apastron X-ray emission primarily magnetospheres of the individual stars, which is supported by the position of X-ray luminosity of $\epsilon$ Lupi during apastron in the XADM X-ray scaling relation. However, for the periastron phase, we considered a combined contribution from binary interaction and single-star emission. Given that the winds of main sequence B-type binary stars are weak, wind-wind collision can be ruled out as a major contributor to binary interaction. Aligning with the interpretation of non-thermal radio emission, this hard X-ray emission can be attributed to magnetospheric interaction (\citealt{Das2023}, this work).

With the superior spectral quality of XMM-Newton, we were able to explore the possible emission mechanisms by fitting different spectral models. We discovered that a power-law model, in combination with two thermal plasma components, can fit the data significantly better than the pure thermal plasma model. The contribution from the power-law component is found to be negligible near apastron, while the contribution becomes most significant during periastron. We thus attribute this non-thermal emission modeled by the power-law to originate from the synchrotron and/or Inverse Compton effect. In an attempt to model the energy released by the magnetosphere, we found that such interactions are indeed capable of accelerating particles to high energies fueling  the observed excess X-rays. As the magnetospheres come very close during periastron, the system undergoes controlled magnetic energy release via the magnetic reconnection process. This phenomenon can give rise to the non-thermal X-ray emission (via synchrotron and/or Inverse Compton) towards periastron similar to the observed synchrotron radio emission. Nevertheless, the lack of knowledge regarding the rotation period of the components hindered our ability to construct a more realistic model.

This study not only complements the findings of \cite{Das2023} by providing data with higher sensitivity and better phase coverage, but also strengthens the scientific understanding of $\epsilon$ Lupi's magnetospheric dynamics through several novel findings:  

\begin{enumerate}
    \item The higher quality data have allowed for more sophisticated modeling, revealing the non-thermal nature of hard X-ray emission, rather than thermal  hard X-ray emission (usually generated in extreme colliding wind scenario). The evidence of non-thermal emission in this system is arguably the most definitive example among magnetic hot stars.

    \item We observed off-periastron enhancements at phases that nearly correspond to those observed at radio bands. This is certainly enlightening and strongly motivates further observations at both radio and X-ray bands.

    \item The X-ray light curve shows a gradual increase of flux towards periastron, suggesting that the periastron enhancement is probably less sharp than what reported by \cite{Das2023}. This finding emphasizes the fueling of X-ray emission through controlled magnetic energy release in the stable magnetospheric configuration of the system.
\end{enumerate}

The observed excess hard X-ray emission during periastron and shorter enhancements during other orbital phases, along with the correlation with radio observations, highlight the intricate nature of binary interactions in massive stellar systems, underscoring the need for further studies to unravel the underlying physical mechanisms. A follow-up study to determine the rotation periods of the components will further refine the model of magnetospheric interaction in this system. Finally, a denser orbital sampling of X-ray emission from this system with sensitive instruments such as XMM-Newton may reveal additional structure in the light curve. Such observations will be crucial to explore additional physical processes in play, if any, that can explain the off-periastron X-ray and radio enhancements.

\begin{acknowledgments}
A.B. and G.A.W. acknowledge support in the form of a Discovery Grant from the Natural Sciences and Engineering Research Council (NSERC) of Canada. This research is based on observations obtained with XMM-Newton, an ESA science mission with instruments and contributions directly funded by ESA Member States and NASA.   
\end{acknowledgments}

%

\vspace{5mm}
\facilities{XMM-Newton. The XMM-Newton data are available from the XMM-Newton Science Archive (XSA, \url{https://www.cosmos.esa.int/web/xmm-newton/xsa}). TESS data is readily available from the Mikulski Archive for Space Telescopes (MAST, \url{https://archive.stsci.edu/}). Analysed data can be obtained from the corresponding author upon request.}

\bibliography{sample631}{}
\bibliographystyle{aasjournal}


\appendix 

\section{mcmc analysis} \label{sec:mcmc}

In this section we show the corner plots from the mcmc analysis for the periastron observation. The analysis was performed using the \textit{chain\footnote{\url{https://heasarc.gsfc.nasa.gov/xanadu/xspec/python/html/chain.html}}} command in PyXspec. The corner plots were made using the package \textit{pyXspecCorner\footnote{\url{https://github.com/garciafederico/pyXspecCorner}}}. For each case, we used chain length of 200000, and a burn length of 2000. The plots for the 4T model are shown in Fig. \ref{fig:A1}, while the plots for 2T+pow model are shown in Fig. \ref{fig:A2}. We do not observe very strong discernible pattern to the distribution, especially in the normalization of the 4.0 keV \textit{apec} component, or the normalization parameter of the \textit{powerlaw} component, which are most relevant in our case. As also seen in the main text, the `nH' parameter is not well fitted in case of 4T model, but relatively well behaved in the 2T+pow model.

\setcounter{figure}{0} 
\makeatletter 
\renewcommand{\thefigure}{A\@arabic\c@figure}
\makeatother

\begin{figure}
    \centering
    \includegraphics[width=0.75\textwidth]{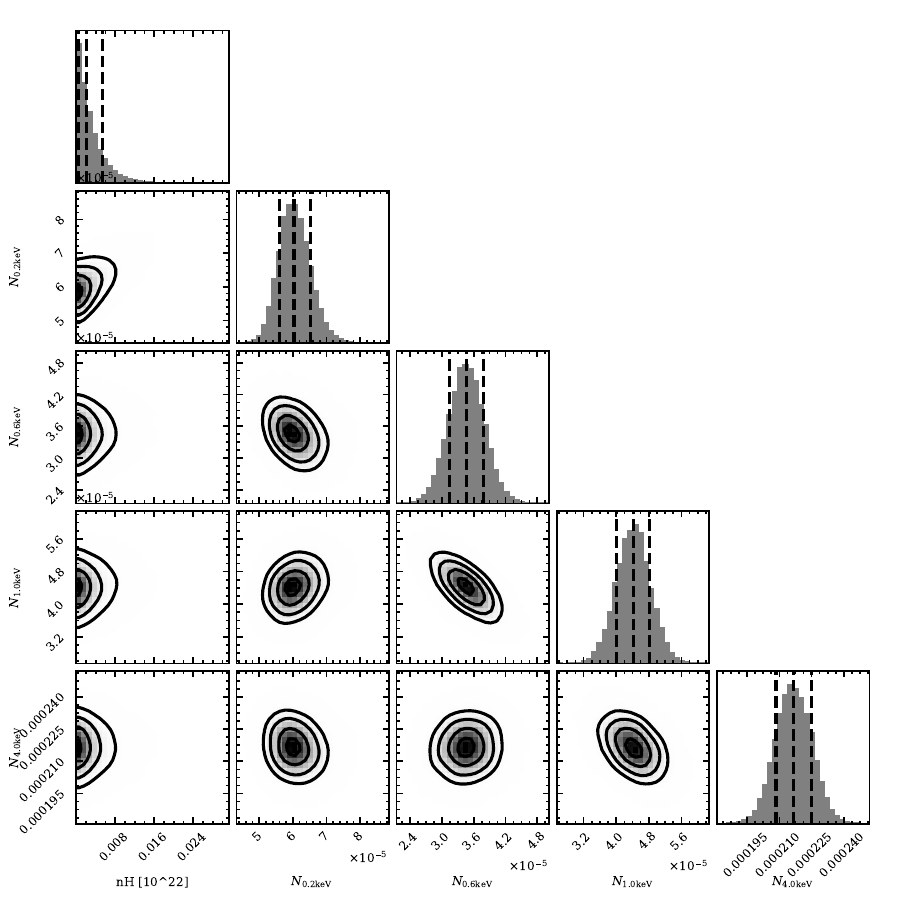}
    \caption{Corner plots corresponding to the 4T fits to the EPIC-pn spectra of periastron observation. $N_{i \ \rm keV}$s ($i=0.2, 0.6, 1.0, 4.0$) are proportional to the normalization factors of the four \textit{apec} models at 0.2, 0.6, 1.0, and 4.0 keV, respectively, in units of $10^{-5}$ cm$^{-5}$). The column density for the \textit{phabs} component is shown with the $n_{\rm H}$ parameter, in units of $10^{22}$ cm$^{-2}$. The dashed vertical lines on the histograms represent the median values (middle), and the 68\% confidence intervals.}
    \label{fig:A1}
\end{figure}

\begin{figure}
    \centering
    \includegraphics[width=0.98\textwidth]{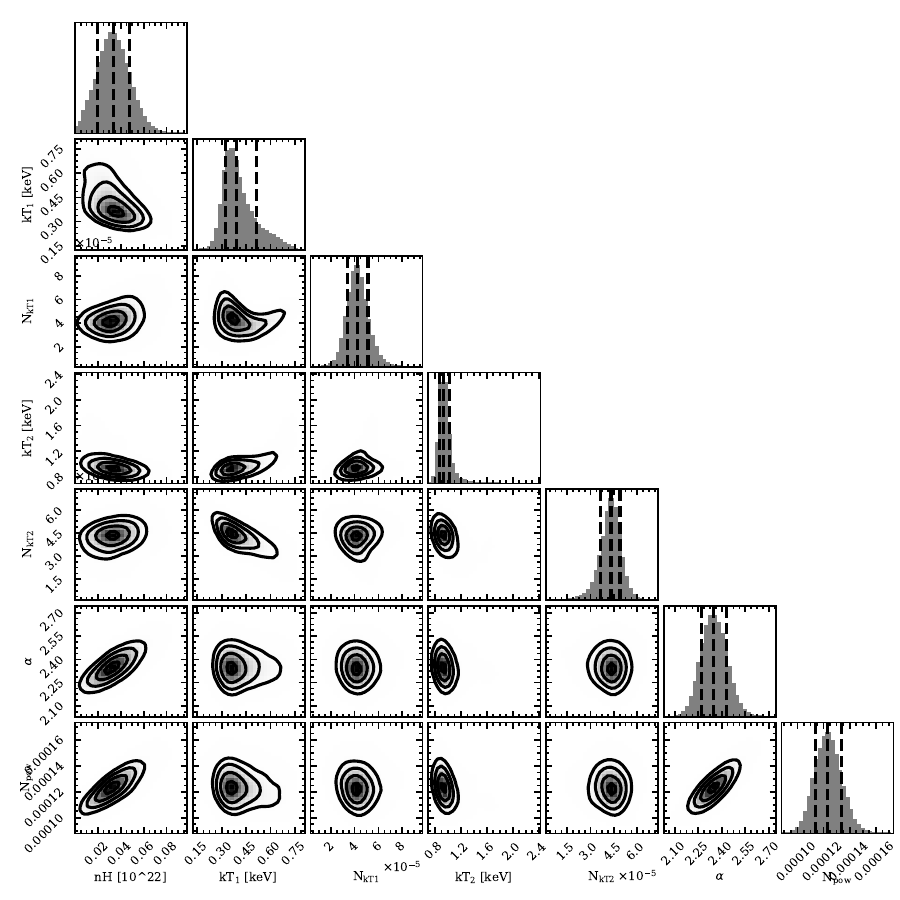}
    \caption{Same as Fig. \ref{fig:A1}, but plots corresponding to the 2T+pow fits to the EPIC-pn spectra of periastron observation.}
    \label{fig:A2}
\end{figure}

\end{document}